\documentclass[twocolumn,fp]{jpsj3}
\usepackage{txfonts}
\usepackage{bm}
\title{Superconductivity and Non-Fermi-Liquid Behavior in the Heavy-Fermion Compound CeCo$_{\bm{1-x}}$Ni$_{\bm x}$In$_{\bm 5}$}

\author{Ryo~Otaka,$^1$ Makoto~Yokoyama,$^1$\thanks{E-mail address: makoto.yokoyama.sci@vc.ibaraki.ac.jp}
Hiroaki~Mashiko,$^1$ Takeshi~Hasegawa,$^1$ Yusei~Shimizu,$^2$ Yoichi~Ikeda,$^3$ Kenichi~Tenya,$^4$ Shota~Nakamura,$^5$ Daichi~Ueta,$^5$ Hideki Yoshizawa,$^5$ and Toshiro~Sakakibara$^5$}
\inst{$^1$Faculty of Science, Ibaraki University, Mito, Ibaraki 310-8512, Japan\\
$^2$Institute for Materials Research, Tohoku University, Oarai, Ibaraki 311-1313, Japan\\
$^3$Institute for Materials Research, Tohoku University, Sendai, Miyagi 980-8577, Japan\\
$^4$Faculty of Education, Shinshu University, Nagano 380-8544, Japan\\
$^5$Institute for Solid State Physics, The University of Tokyo, Kashiwa, Chiba 277-8581, Japan
} 

\abst{
The effect of off-plane impurity on superconductivity and non-Fermi-liquid (NFL) behavior in the layered heavy-fermion compound CeCo$_{1-x}$Ni$_x$In$_5$ is investigated by specific heat, magnetization, and electrical resistivity measurements. These measurements reveal that the superconducting (SC) transition temperature $T_{\rm c}$ monotonically decreases from 2.3 K ($x=0$) to 0.8 K ($x=0.20$) with increasing $x$, and then the SC order disappears above $x=0.25$. At the same time, the Ni substitution yields the NFL behavior at zero field for $x=0.25$, characterized by the $-\ln T$ divergence in specific heat divided by temperature, $C_p/T$, and magnetic susceptibility, $M/B$. The NFL behavior in magnetic fields for $x=0.25$ is quite similar to that seen at around the SC upper critical field in pure CeCoIn$_5$, suggesting that both compounds are governed by the same antiferromagnetic quantum criticality. The resemblance of the doping effect on the SC order among Ni- , Sn-, and Pt-substituted CeCoIn$_5$ supports the argument that the doped carriers are primarily responsible for the breakdown of the SC order. The present investigation further reveals the quantitative differences in the trends of the suppression of superconductivity between Ce(Co,Ni)In$_5$ and the other alloys, such as the rates of decrease in $T_{\rm c}$, $dT_{\rm c}/d x$, and specific heat jump at $T_{\rm c}$, $d(\Delta C_p/T_{\rm c})/d x$. We suggest that the occupied positions of the doped ions play an important role in the origin of these differences. 
}

\begin{document}
\maketitle
\section{Introduction}
The relationship between unconventional superconductivity (SC) and quantum critical phenomena has been extensively investigated in the physics of heavy-fermion systems. In particular, Ce-based heavy-fermion superconductors have been continuously spotlighted since in most of these compounds, the SC phase emerges at the antiferromagnetic (AFM) quantum critical point (QCP) \cite{rf:Mathur98}. The AFM-QCP, corresponding to the AFM transition at zero temperature, is often generated by suppressing the AFM order via applying pressure, magnetic field, and chemical substitutions. Various macroscopic quantities, such as specific heat, magnetic susceptibility, and electrical resistivity, in the vicinity of the AFM-QCP exhibit the non-Fermi-liquid (NFL) behavior originating from the strong enhancement of the AFM quantum critical fluctuations \cite{rf:Pfleiderer2009,rf:Stewart2001,rf:Lohneysen2007}. It is therefore expected that this fluctuation is tightly coupled with the Cooper pairing in these compounds. 

Among the Ce-based heavy-fermion superconductors, CeCoIn$_5$ shows particularly intriguing properties concerning the SC and the AFM quantum critical phenomena. This compound has a primitive tetragonal crystal structure (the HoCoGa$_5$-type structure) composed of the stacking sequence of CeIn--In$_2$--Co--In$_2$ layers along the $c$-axis [inset of Fig.\ 1(b)], and hence, two-dimensional characteristics of the conduction electrons are expected \cite{rf:Hall2001}. The SC transition of CeCoIn$_5$ at $T_{\rm c}=2.3\ {\rm K}$ is characterized by an anomalously large specific heat jump of $\Delta C/\gamma T_{\rm c}=4.5$ in stark contrast to that of the weak coupling BCS value (1.43) \cite{rf:Petrovic2001}. The magnetic origin of the SC pairing is inferred from the $d$-wave ($d_{x^2-y^2}$) symmetry of the SC gap determined by thermal conductivity, specific heat, and conductance measurements \cite{rf:Izawa2001,rf:An2010,rf:Park2008}. In magnetic fields, the spin degrees of freedom are significantly coupled with the stability of the SC order; a strong Pauli paramagnetic effect gives rise to a first-order transition at the SC upper critical field $H_{\rm c2}$ below 0.7 K \cite{rf:Izawa2001,rf:Tayama2002,rf:Ikeda2001,rf:Bianchi2002}, and the SC phase coexistent with AFM spin modulation evolves just below $H_{\rm c2}$ at very low temperatures \cite{rf:Bianchi2003-1,rf:Radovan2003,rf:Kakuyanagi2005,rf:Young2007,rf:Kenzelmann2008,rf:Tokiwa2008}. Furthermore, the existence of the AFM-QCP at $\sim H_{\rm c2}$ is strongly suggested from the observations of the NFL behavior in the paramagnetic phase above $H_{\rm c2}$, including the $-\ln T$ divergence in specific heat divided by temperature, the $T$-linear dependence in magnetization, and electrical resistivity \cite{rf:Tayama2002,rf:Bianchi2003-2,rf:Paglione2003}. In fact, the long-range AFM orders are generated by substituting the ions for the elements in CeCoIn$_5$, such as Nd for Ce \cite{rf:Hu2008,rf:Raymond2014}, Rh for Co \cite{rf:Zapf2001,rf:Yoko2006,rf:Yoko2008,rf:Ohira-Kawamura2007}, and Cd, Hg, and Zn for In \cite{rf:Pham2006,rf:Nicklas2007,rf:Yoko2014,rf:Yoko2015}. 

In contrast to the doping effect by those ions, the substitution of Sn for In simply suppresses the SC phase without generating the AFM order \cite{rf:Bauer2005,rf:Bauer2006,rf:Ramos2010}. Both $T_{\rm c}$ and $H_{\rm c2}$ are monotonically reduced by doping Sn, and then become zero at the Sn concentration of 18\%. At this critical Sn concentration, the NFL behavior is realized at zero magnetic field, corresponding to the $T$-linear dependence in electrical resistivity and the $-\ln T$ divergence in specific heat divided by temperature. The similarity of the NFL behaviors between CeCo(In,Sn)$_5$ and pure CeCoIn$_5$ indicates that the AFM quantum critical fluctuation is still enhanced in CeCo(In,Sn)$_5$, whereas no AFM ordering is observed in a wide Sn concentration range. Furthermore, it is suggested that these features would be related to the two-dimensional nature, since the X-ray absorption study revealed that most of the doped Sn ions occupy the CeIn layer located at the tetragonal basal plane \cite{rf:Daniel2005}. In contrast to this suggestion, it is argued from the similarity in the suppression of the SC order between Ce(Co,Pt)In$_5$ and CeCo(In,Sn)$_5$ that the effect of the impurity on the SC order is independent of the occupied impurity positions and layers \cite{rf:Gofryk2012}. The suppression of the SC order is also observed in pure CeCoIn$_5$ under pressure \cite{rf:Sidorov2002}, but the usual Fermi-liquid state instead of the NFL state is stabilized at the critical pressure of the SC order. 

CeCoIn$_5$ is thus suggested to show the close relationship among the SC order, the AFM quantum critical fluctuation, and the dimensionality of the conduction electrons, but the entire physical properties are still unclear. To obtain a comprehensive understanding of the relationship among them, we have investigated the low-temperature properties of the new doped alloys CeCo$_{1-x}$Ni$_x$In$_5$ for the first time. In particular, we have succeeded in tuning the suppression of the SC phase without the AFM ordering, by replacing the Co ions located out of the CeIn layers. Since both the Ni and Sn ions have the neighbor atomic numbers of the Co and In ions, respectively, the investigation of the Ni doping is useful for comparing the SC-breaking features and the NFL anomaly between the off-plane impurity doping [Ce(Co,Ni)In$_5$] and the in-plane impurity doping [CeCo(In,Sn)$_5$]. In this paper, we describe the properties of the SC order and the NFL behaviors in the Ni-doped CeCoIn$_5$ investigated by magnetization, specific heat, and electrical resistivity measurements, and discuss the experimental results in comparison with those observed in the other doped alloys. 

\section{Experimental Procedure}
Plate-shaped single crystals of CeCo$_{1-x}$Ni$_x$In$_5$ with $x \le 0.3$, whose basal planes are perpendicular to the tetragonal $c$-axis, were grown by the indium flux technique. Appropriate amounts of Ce, Co, Ni, and In with an excess flux were set in quartz tubes, and then sealed under 0.02 MPa Ar atmosphere. They were heated up to 1050 $^\circ$C and then cooled by a two-stage process similar to the previously reported one \cite{rf:Petrovic2001}. The tetragonal structure of the obtained single crystals was checked by the X-ray diffraction. The energy dispersive X-ray spectroscopy (EDS) measurements for these crystals indicate the homogeneous distributions of all the elements, and the estimated Ni/Co concentrations approximately coincide with the starting (nominal) values. To further check the Ni/Co concentration, we performed the inductively coupled plasma mass spectrometry (ICP-MS) measurements for pieces picked up from the obtained single crystals, and found that the correspondence between the nominal and the measured $x$ values is achieved within the deviation of $\Delta x/x\sim 17\%$ including the experimental error. Hereafter, we use the nominal $x$ values, along with the error of $x$ estimated from these analyses.

The electrical resistivity $\rho$ was measured by a standard four-wire technique, and specific heat $C_p$ measurements were carried out by the relaxation method. Both measurements were performed at temperatures down to 0.5 K using a commercial measurement system (PPMS, Quantum Design) and handmade equipment. In the $\rho$ measurements, an external magnetic field $B\ (=\mu_0H)$ ($\mu_0$: vacuum permeability) of 0--7 T was applied perpendicular to the sample current $j$. The magnetization $M$ was measured using a capacitively detected Faraday-force magnetometer \cite{rf:Sakakibara94} between 0.27 and 2.2 K and in fields up to 8.5 T. A commercial SQUID magnetometer (MPMS, Quantum Design) was also used for magnetization measurements in the temperature range of 2.0--300 K and fields up to 5 T. The $a$-axis ac susceptibility $\chi_{\rm ac}$ was measured between 1.1 and 4 K by a standard Hartshorn-bridge method, in which an ac field is applied parallel to the basal plane of the plate-shaped samples in order to reduce the demagnetizing-field effect. The magnitude and frequency of the applied ac field were selected to be 0.05 mT and 180 Hz, respectively.

\section{Results}
\subsection{Physical properties at zero and weak magnetic fields}
\begin{figure}[tbp]
\begin{center}
\includegraphics[bb=40 151 443 819,keepaspectratio,width=0.42\textwidth]{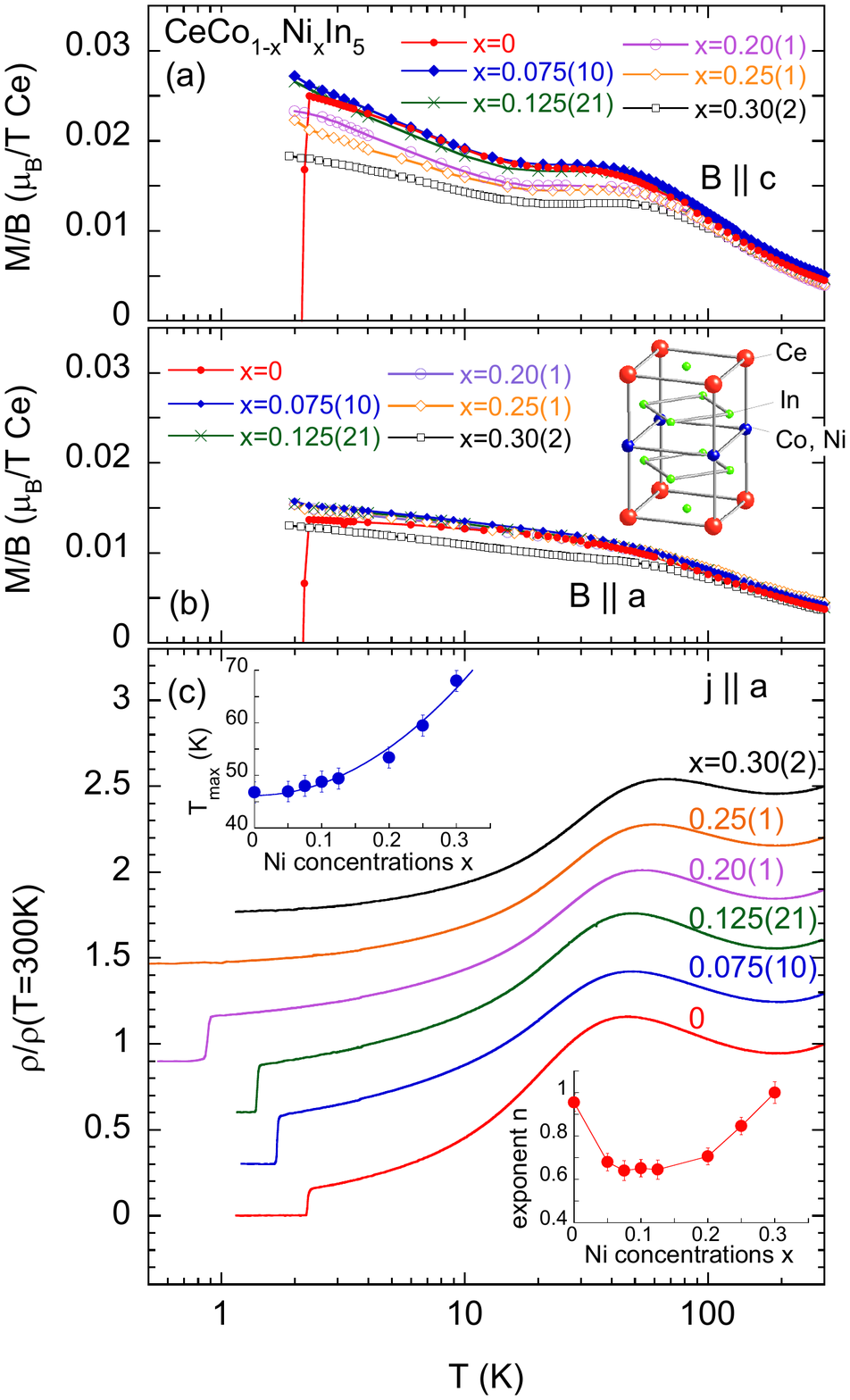}
\end{center}
  \caption{
(Color online)  Temperature variations of the magnetization divided by magnetic field, $M/B$, for (a) $B\,||\,c$ and (b) $B\,||\,a$, and (c) the $a$-axis electrical resistivity $\rho$ normalized by the magnitude at 300 K for CeCo$_{1-x}$Ni$_x$In$_5$. In (a) and (b), the applied field ranges from 0.1 to 0.5 T. In (c), the vertical baselines of the $\rho$ data for $x\ge 0.075$ are shifted in steps of 0.3 for clarity. The crystal structure of Ce(Co,Ni)In$_5$ is depicted in the inset of (b). Shown in the upper and lower insets of (c) are the Ni concentration dependences of $T_{\rm max}$ and the exponent $n$ of the $T^n$ function in $\rho(T)$, respectively. The solid curve in the former inset indicates the $x^2$ fit for $T_{\rm max}$. The exponents $n$ in the latter inset of (c) are obtained by fitting the $\rho(T)$ data between 2.5 and 15 K.
}
\end{figure}
Figures 1(a) and 1(b) show temperature variations of the $c$- and $a$-axes magnetic susceptibility $M/B$, respectively, obtained by the SQUID magnetometer. $M/B$ for the $c$-direction shows a shoulder-like behavior, whose temperature increases from $\sim 45\ {\rm K}$ ($x=0$) to $\sim 60\ {\rm K}$ [$x=0.25(1)$] with increasing $x$. Such a trend is also seen in the $a$-axis $M/B$ data, but it is less clear than that of the $c$-axis. The $M/B$ curves above these temperatures fairly well follow the Curie$-$Weiss law described by $N\mu_{\rm eff}^2/3k_{\rm B}(T-\theta_{\rm p})$ for all the Ni concentrations presently investigated. The best fits for the $M/B$ data in the temperature range of 120--300 K at $x=0.125(21)$ give the $\mu_{\rm eff}$ and $\theta_{\rm p}$ values of 2.6(1) $\mu_{\rm B}/{\rm Ce}$ and $-35(5)$ K for the $c$-axis, and 2.5(1) $\mu_{\rm B}/{\rm Ce}$ and $-87(5)$ K for the $a$-axis, respectively, and similar values are obtained in the entire Ni concentration range. These $\mu_{\rm eff}$ values are comparable to 2.54 $\mu_{\rm B}$ calculated from the $J=5/2$ multiplet in the Ce$^{3+}$ ion. The negative $\theta_{\rm p}$ values indicate that the AFM correlation is dominant for all the Ni-doped alloys. At low temperatures, anisotropic variations in the magnitude of $M/B$ occur as $x$ is increased; doping Ni gradually reduces the $c$-axis $M/B$ values while yielding a slight change in the $a$-axis ones. A drop of $M/B$ at 2.3 K for pure CeCoIn$_5$ reflects the shielding due to the SC order. This effect is invisible for the Ni-doped samples because $T_{\rm c}$ becomes lower than the lowest accessible temperature (2.0 K) of the SQUID magnetometer.

Displayed in Fig.\ 1(c) are temperature variations of the $a$-axis electrical resistivity $\rho(T)$ for CeCo$_{1-x}$Ni$_x$In$_5$, normalized by the magnitude at 300 K. In the present experiments, the shape and the properties of the experimentally obtained $\rho(T)$ function at each $x$ are well reproduced, but the magnitudes of $\rho$ are somewhat sample-dependent, ranging from 40 to 100 $\mu\Omega$ cm at 300 K. The SC transition, characterized by the drop in $\rho(T)$, is monotonically suppressed from $T_{\rm c}=2.25\ {\rm K}$ ($x=0$) to 0.88 K [$x=0.20(1)$] by doping Ni, and then disappears for $x\ge 0.25$. Above $T_{\rm c}$, $\rho(T)$ increases with increasing temperature, and then shows a broad peak at $T_{\rm max}$ ranging from 46 K ($x=0$) to 68 K [$x=0.30(2)$], where $M/B$ exhibits the shoulder-like behavior. $T_{\rm max}$ is found to increase with increasing $x$, suggesting that the energy scale of the coherent heavy-fermion state becomes large with doping Ni. The increase in $T_{\rm max}$ is similar to those observed in Sn- and Pt-doped CeCoIn$_5$, but different from the decrease in $T_{\rm max}$ found in the Cd, Hg, and Zn substitutions \cite{rf:Yoko2014,rf:Bauer2006,rf:Gofryk2012}. These opposite variations would mainly be attributed to the sign of the doped charge carrier in these compounds, as pointed out previously \cite{rf:Gofryk2012}. In the Ni-doped alloys, however, $T_{\rm max}$ is found to vary in proportion to $x^2$ [upper inset of Fig.\ 1(c)], showing a remarkable discrepancy from the global $x$-linear variation of $T_{\rm max}$ indicated in X-doped CeCoIn$_5$ (X=Cd, Hg, Sn, and Pt) \cite{rf:Gofryk2012}. 

It is known that $\rho(T)$ between $T_{\rm c}$ and $T_{\rm max}$ for CeCoIn$_5$ exhibits the NFL behavior with the $T$-linear dependence \cite{rf:Petrovic2001,rf:Paglione2003}. A similar NFL behavior is also seen in the Ni substitutions. The exponent $n$ of the $T^n$ function in $\rho(T)$, obtained by simply fitting the $\rho(T)$ data between 2.5 and 15 K, decreases to 0.65(5) at $x=0.125$, but then recovers to 0.85(4) at $x=0.25$ and 1.00(5) at $x=0.30(2)$ [lower inset of Fig.\ 1(c)]. Quite similar values of $n$ are obtained for $x \ge 0.25$ even when the lower limit of the fitting range is extended to the lowest accessible temperature in the present experiments. It is unclear in the present stage why the $x$ variations of $n$ have a minimum at $x\sim 0.1$. It appears that such a broad minimum of $n$ does not come from the magnetic quantum criticality. A possible reason may be the competing effects that $n$ decreases and increases with increasing $x$; the former is often observed in the other doped CeCoIn$_5$ \cite{rf:Yoko2014,rf:Paglione2007} although its origin is still unclear, and the latter would be expected when doping a large amount of Ni favors the Fermi-liquid state.

In Figs.\ 2(a)--2(c), we plot the $a$-axis volume ac susceptibility, $\chi_{\rm ac}$, specific heat divided by temperature, $C_p/T$, and electrical resistivity, respectively, in the vicinity of $T_{\rm c}$. For $x\le 0.125$, the SC transition at each $x$ is recognized by a large and discontinuous diamagnetic signal in $\chi_{\rm ac}$, a clear jump in $C_p/T$, and a drop of $\rho$ toward zero at $T_{\rm c}$. The magnitudes of the diamagnetic signal in $\chi_{\rm ac}$ are comparable among all the samples belonging to this $x$ range, indicative of the full-volume SC order being achieved below $T_{\rm c}$. At the same time, the magnitude of the specific heat jump $\Delta C_p/T_{\rm c}$ associated with the SC transition is found to be gradually reduced from 1.6 J/K$^2$ mol ($x=0$) to 1.2 J/K$^2$ mol ($x=0.125$) with increasing $x$. At $x=0.20$, by contrast, the $C_p/T$ jump becomes broad and its magnitude markedly decreases to 0.34 J/K$^2$ mol. This sudden drop in the $\Delta C_p/T_{\rm c}$ value may indicate that the SC volume fraction is reduced as $T_{\rm c}$ becomes close to zero. Unfortunately, however, we cannot estimate the magnitude of the shielding signal in $\chi_{\rm ac}$ at $x=0.20$ since $T_{\rm c}$ ($=0.88\ {\rm K}$) is lower than the lowest accessible temperature (1.1 K) of our $\chi_{\rm ac}$ experiment. With further doping Ni, our magnetization measurements detect no signature associated with the phase transition including the SC order down to 0.27 K, and it is therefore expected that the SC order becomes fully suppressed and the paramagnetic ground state becomes stable for $x\ge 0.25$. Interestingly, $C_p/T$ at $x=0.25$ shows the NFL behavior with the $-\ln T$ divergence, and it is rather suppressed at $x=0.30$ [inset of Fig.\ 2(b)]. The NFL behaviors presently observed in $C_p/T$ and $\rho$ at $x=0.25$ are quite similar to those found at $\sim H_{\rm c2}$ in CeCoIn$_5$ as well as at the critical Sn concentration in CeCo(In,Sn)$_5$ \cite{rf:Bianchi2003-2,rf:Bauer2005,rf:Bauer2006}. 

\begin{figure}[tbp]
\begin{center}
\includegraphics[bb=21 103 384 819,keepaspectratio,width=0.42\textwidth]{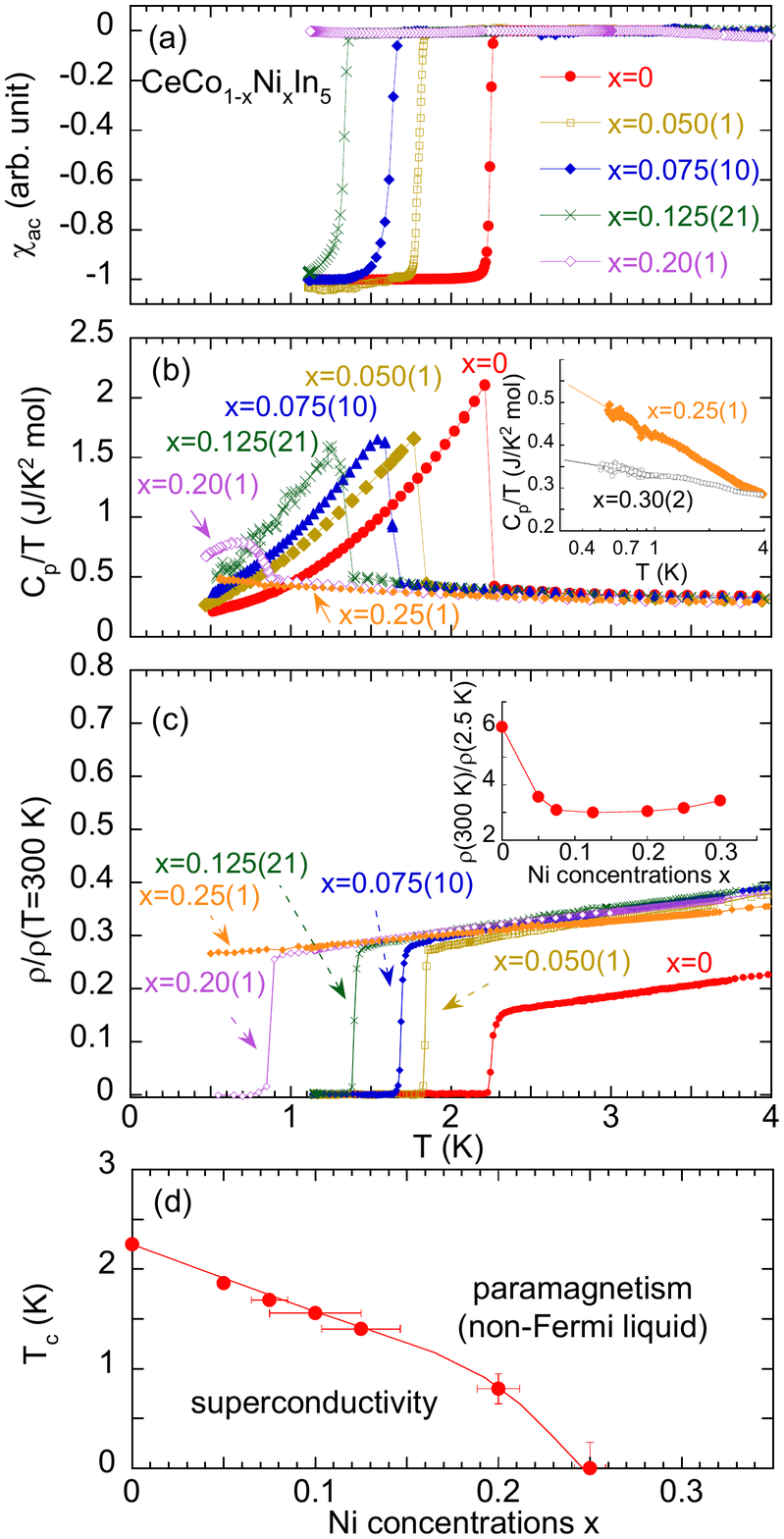}
\end{center}
  \caption{
(Color online)  Temperature variations of (a) $a$-axis volume ac susceptibility $\chi_{\rm ac}$, (b) specific heat divided by temperature, $C_p/T$, and (c) electrical resistivity normalized by the magnitude at 300 K for CeCo$_{1-x}$Ni$_x$In$_5$, plotted in the vicinity of the SC transition. The inset of (b) is the $\log T$ plot of $C_p/T$ for $x=0.25$ and 0.30, and the inset of (c) shows $x$ variations of the residual resistivity ratio defined by $\rho(300\ {\rm K})/\rho(2.5\ {\rm K})$. (d) $x-T$ phase diagram of CeCo$_{1-x}$Ni$_x$In$_5$ obtained from temperature variations of ac susceptibility, magnetization, electrical resistivity, and specific heat. The line in (d) is a guide to the eye.
}
\end{figure}

In Fig.\ 2(d), we summarize the $x-T$ phase diagram at zero field for CeCo$_{1-x}$Ni$_x$In$_5$. $T_{\rm c}$ shows a linear decrease with increasing $x$ up to 0.20, whose rate $d T_{\rm c}/d x$ is estimated to be $-7.1(4)$ K/$x$(Ni). The $x$-linear decrease in $T_{\rm c}$ is also found in the Sn- and Pt-doped alloys, and this similarity suggests that a common mechanism on the breakdown of the SC order exists in these doped alloys, as suggested previously \cite{rf:Gofryk2012}. From a quantitative point of view, however, the decreasing rate for the Ni-doped alloys is significantly smaller than those for CeCoIn$_{5-x}$Sn$_x$ [$-12$ K/$x$(Sn)] \cite{rf:Bauer2005} and CeCo$_{1-x}$Pt$_x$In$_5$ [$-11$ K/$x$(Pt)] \cite{rf:Gofryk2012}. These results indicate that the doped elements clearly differ in their ability to disturb the SC order. In the next section, we will discuss the difference in further detail.

\subsection{Superconducting properties in magnetic fields}
Figure 3 shows the field variations of the magnetization $M(B)$ at 0.35 K for CeCo$_{1-x}$Ni$_x$In$_5$, obtained using the capacitive Faraday-force magnetometer. The equilibrium magnetization curves $M_{\rm eq}(B)$, which are estimated from an average of the $M(B)$ data taken under increasing and decreasing field variations, are also indicated as the solid lines in Fig.\ 3. The hysteretic behavior in $M(B)$ for $x \le 0.20$ is indicative of the existence of the SC order. The very small hysteresis loop in $M(B)$ for pure CeCoIn$_5$ reflects the high quality of the sample, since such an irreversible $M(B)$ feature in the SC region is usually caused by the flux pinning effect, which generally occurs on impurities and lattice defects. Thus, the enlarged hysteresis loop with increasing $x$ up to 0.125 would be ascribed to the disorder effect generated by doping. This effect is also considered to generate the steep decrease in the residual resistivity ratio $\rho(300\ {\rm K})/\rho(2.5\ {\rm K})$ with doping Ni [inset of Fig.\ 2(c)]. However, the loop widths in $M(B)$ for the Ni-doped alloys are still much ($\sim 5-40$ times) smaller than those observed in the $c$-axis $M(B)$ curve of CeCo(In,Zn)$_5$ with nearly the same amount of Zn substitutions \cite{rf:Yoko2015}. This is considered to be caused by the difference in the assigned positions of the doped ions in the crystal, as will be argued in the next section. At $x=0.20$, the irreversible feature in the $M(B)$ data as well as the anomaly at $H_{\rm c2}$ in $M_{\rm eq}(B)$ are weaker than those for $x =0.050(1)$ and 0.125 (inset of Fig.\ 3). This may be related to the shrinkage of the $\Delta C_p/T_{\rm c}$ value at the corresponding Ni concentration. For $x\ge 0.25$, no hysteretic behavior was observed in the magnetization data down to 0.27 K.

The breakdown of the SC order at the upper critical field $H_{\rm c2}$ can be recognized by the jump or kink anomaly in $M_{\rm eq}(B)$ as well as the closing of the hysteresis loop in $M(B)$. As reported previously \cite{rf:Tayama2002}, a first-order nature of the SC-to-paramagnetic transition is evidenced by a clear discontinuous jump of $M_{\rm eq}$ at $\mu_0H_{c2}=4.85\ {\rm T}$. Doping Ni into CeCoIn$_5$ monotonically reduces the magnitudes of $\mu_{\rm 0}H_{\rm c2}$ to 3.75(5) T at $x=0.050$ and 1.43(5) T at $x=0.20$, and this reduction is consistent with the $x$-linear suppression of $T_{\rm c}$ for $x\le 0.20$. At $x=0.050$ and 0.125, a peak effect seen in $M(B)$ obscures the thermodynamic features in $M(B)$ concerning the breakdown of the SC. Nevertheless, $M_{\rm eq}(B)$ seems to change continuously at $H_{\rm c2}$, indicating that the first-order transition of the SC breaking by the magnetic field changes into the second-order one with increasing $x$. This variation is also found in the other doped alloys \cite{rf:Tokiwa2008,rf:Yoko2014}, and hence, the disorder induced by doping is considered to commonly affect the nature of the transition at $H_{\rm c2}$. 
\begin{figure}[tbp]
\begin{center}
\includegraphics[bb=41 204 469 670,keepaspectratio,width=0.42\textwidth]{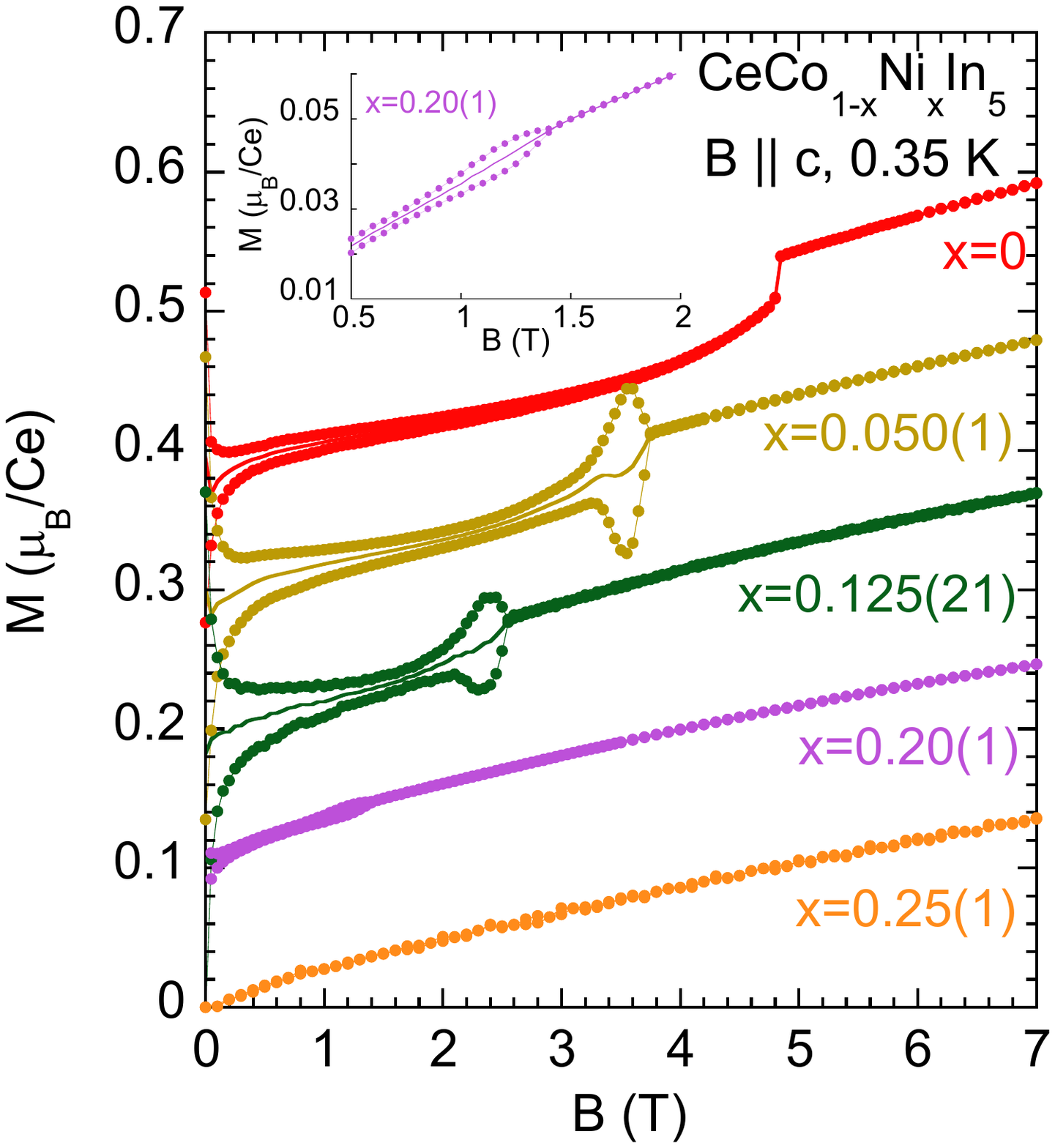}
\end{center}
  \caption{
(Color online)   Magnetization curves at 0.35 K for CeCo$_{1-x}$Ni$_x$In$_5$, measured under the increasing and decreasing fields for $B\,||\,c$. Note that the offsets of the magnetization for $x\le 0.20$ are shifted in steps of 0.1 $\mu_{\rm B}/{\rm Ce}$ for clarity. Solid lines are the equilibrium magnetization curves estimated from an average of the hysteresis loops. Displayed in the inset is the enlarged magnetization curve around $\mu_{\rm 0}H_{\rm c2}$ for $x=0.20$.
}
\end{figure}

In Fig.\ 4, we plot the temperature dependence of the $c$-axis upper critical field $H_{\rm c2}(T)$ for CeCo$_{1-x}$Ni$_x$In$_5$, estimated from the electrical resistivity and magnetization measurements under magnetic fields. The evaluated SC boundaries for $x \le 0.20$, including that for pure CeCoIn$_5$ \cite{rf:Tayama2002}, are quite analogous to each other, strongly suggesting that Ni doping simply contracts the SC condensation energy. This feature is also found in the CeCo(In,Sn)$_5$ alloys \cite{rf:Ramos2010}, but it is quite different from the unusual $x$ variations of $H_{\rm c2}$ revealed in CeCo(In,Cd)$_5$ \cite{rf:Tokiwa2008} and CeCo(In,Zn)$_5$ \cite{rf:Yoko2015}, in which doping Cd or Zn in CeCoIn$_5$ slightly increases the low-temperature $H_{\rm c2}$, while markedly decreasing $T_{\rm c}$. We have suggested that such $H_{\rm c2}$ behavior in CeCo(In,Zn)$_5$ arises from the relaxation of the Pauli paramagnetic effect yielded by evolving the AFM correlation \cite{rf:Yoko2015}. In this context, we expect that the SC order in the Ni-doped alloys is still governed by the Pauli-limited condition. To verify this expectation, we estimate the orbital-limited critical field at zero temperature $H_{\rm c2}^{\rm orb}(0)$ by using a simple relation: $\mu_0H_{c2}^{\rm orb}(0) \sim -0.7\, T_c\, (\mu_0dH_{c2}/dT)_{T_c}$ \cite{rf:Werthammer66}. It is found that the initial slope of $H_{\rm c2}(T)$, $(\mu_0dH_{c2}/dT)_{T_c}$, is roughly independent of $x$ for $x\le 0.20$ (inset of Fig.\ 4), yielding the ratio $H_{\rm c2}^{\rm orb}(0)/H_{\rm c2}(0) \sim 3$ for all these alloys. This coincidence of the ratio among the pure and Ni-doped compounds naturally leads to the conclusion that the Pauli-limited condition is unchanged upon doping of Ni into CeCoIn$_5$. As a consequence, the Maki parameter $\alpha=\sqrt{2} H_{\rm c2}^{\rm orb}/H_{\rm P}$ ($H_{\rm P}$: the Pauli-limited field) of these alloys is deduced to be $\sim 4$ by assuming that the condition $H_{\rm c2}\sim H_{\rm P}$ holds at low temperatures.
\begin{figure}[tbp]
\begin{center}
\includegraphics[bb=46 163 511 626,keepaspectratio,width=0.42\textwidth]{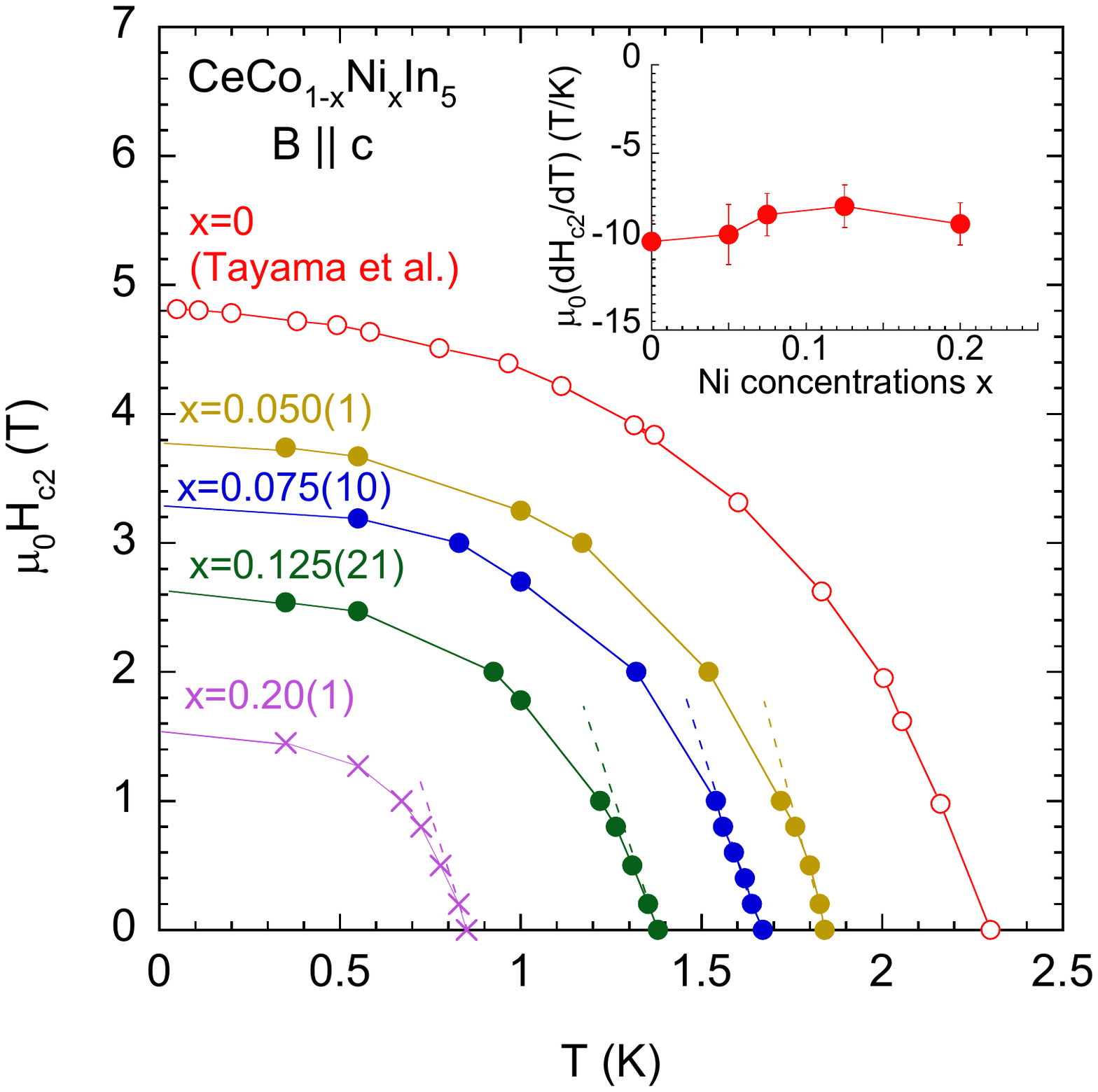}
\end{center}
  \caption{
(Color online)  Temperature variations of the $c$-axis upper critical field $\mu_0H_{\rm c2}(T)$ for CeCo$_{1-x}$Ni$_x$In$_5$ ($x=0.050$, 0.075, 0.125, and 0.20), obtained from the temperature and field variations of the electrical resistivity and magnetization. The $\mu_0H_{\rm c2}(T)$ curve for pure CeCoIn$_5$ \cite{rf:Tayama2002} is also plotted for comparison. The broken lines indicate the slope of $\mu_0H_{\rm c2}(T)$ at $\sim T_{\rm c}$. The inset shows the $x$ variations of the slope of $\mu_0H_{\rm c2}(T)$ at $\sim T_{\rm c}$, in which the value at $x=0$ is taken from Ref. 30.
}
\end{figure}

\subsection{Comparison of the non-Fermi liquid behavior between pure and Ni-doped CeCoIn$_5$}
At $x=0.25$, we have observed the NFL behavior characterized by the $-\ln T$ divergence in $C_p/T$ at zero field. This is quite similar to the observation at $\sim 5\ {\rm T}$ ($\sim \mu_0H_{\rm c2}$) in pure CeCoIn$_5$ \cite{rf:Bianchi2003-2}. To obtain further information on the similarity or difference in the NFL anomaly between them, we have measured the magnetization in these NFL regions. In Fig. 5, we compare the temperature variations of $M/B$ along the $c$-axis for $x=0.25$ and $x=0$. As is seen in Fig.\ 5(a), $M/B$ at 0.5 T for $x=0.25$ exhibits a clear $-\ln T$ divergence in a wide temperature range below $\sim 10\ {\rm K}$. This behavior is different from the slightly concave-upward curve of $M/B$ at 5 T for $x=0$ in the linear temperature scale [Fig.\ 5(b) and its inset]. Moreover, applying a magnetic field easily suppresses the diverging behavior in $M/B$ for the $x=0.25$ sample. The $M/B$ curves at 1--2 T show an almost $T$-linear increase [inset of Fig.\ 5(a)], and are comparable to that at 5 T in $x=0$, possibly reflecting that these NFL behaviors arise from the same mechanism. 
\begin{figure}[tbp]
\begin{center}
\includegraphics[bb=46 294 507 721,keepaspectratio,width=0.42\textwidth]{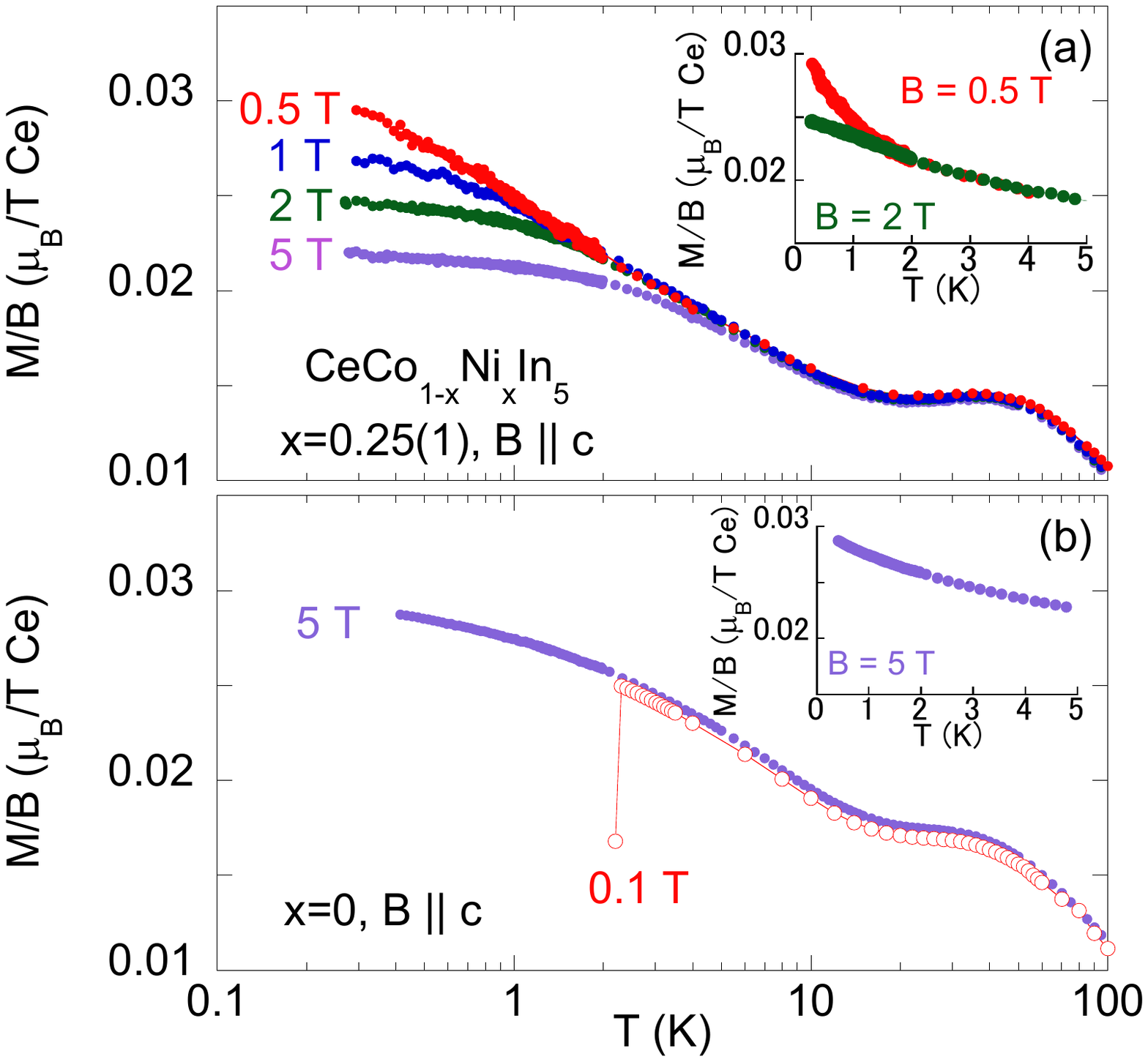}
\end{center}
  \caption{
(Color online) ${\rm Log} T$ plots of the $c$-axis magnetization divided by magnetic field, $M/B$, for CeCo$_{1-x}$Ni$_x$In$_5$ with (a) $x=0.25$ and (b) $x=0$. The insets show the $M/B$ data plotted with the linear temperature scale.
}
\end{figure}
\section{Discussion}
\subsection{Relationship between the doped ionic position and the suppression of the superconductivity}
It has been suggested that the doping of the electron (Pt and Sn) and hole (Hg and Cd) for CeCoIn$_5$ can systematically tune the suppression of the superconductivity; the former stabilizes the NFL paramagnetic state, and the latter induces the AFM order \cite{rf:Gofryk2012}. In this study, we have clarified that the overall features on the suppression of the superconductivity in the Ni-doped alloys are quite similar to those revealed in CeCo(In,Sn)$_5$ \cite{rf:Bauer2005,rf:Bauer2006} and Ce(Co,Pt)In$_5$ \cite{rf:Gofryk2012}. It is thus considered that our present investigation adds another example of the reduction of the SC order parameter caused by the increase in the number of electron carriers by doping. However, the present experimental results also provide the quantitative differences in the SC and paramagnetic properties between the Ni-doped alloys and the others. The Ni-doped alloys show the $T_{\rm max} \propto x^2$ behavior contrary to the $x$-linear variations of $T_{\rm max}$ observed in the other doped alloys. Furthermore, the decreasing rate of $T_{\rm c}$ with $x$ in Ce(Co,Ni)In$_5$ is roughly 60\% of those in the Sn- and Pt-doped alloys. These differences indicate that there is some active contribution to the variations of $T_{\rm c}$ and $T_{\rm max}$, along with the number and sign of the doped carrier \cite{rf:Pt-and-Ni}. 

In particular, it should be stressed that the occupied crystallographic positions of the doped ions would significantly be coupled with the SC properties. Note that in Sn- or Zn-doped CeCoIn$_5$, a considerable amount of the doped Sn or Zn ions fits in the CeIn layer, whereas in Ce(Co,Ni)In$_5$, the doped Ni ions do not enter there. One remarkable feature presumably related to the doped ionic positions is that the decreasing rate of $\Delta C_p/T_{\rm c}$ with $x$, $d(\Delta C_p/T_{\rm c})/d x$, for $x \le 0.125$ in CeCo$_{1-x}$Ni$_x$In$_5$ is roughly one-third of that in CeCoIn$_{5-x}$Sn$_x$ \cite{rf:Bauer2006}. It is suggested in CeCo(In,Sn)$_5$ that the unitarity-limit scattering by the nonmagnetic Sn impurity is responsible for the suppression of $\Delta C_p/T_{\rm c}$ \cite{rf:Bauer2006}. In this context, we expect that the doped Ni ions lead to weaker impurity scattering than the Sn ions when the doped Ni ions are also regarded as nearly nonmagnetic impurities in accordance with the discussion given for the high-$T_{\rm c}$ superconductor \cite{rf:Hudson2001}. In fact, the values of $\rho(300\ {\rm K})/\rho(2.5\ {\rm K})$ for CeCo$_{1-x}$Ni$_x$In$_5$ with $x \le 0.2$ [inset of Fig. 2(c)] are roughly $1.2-1.25$ times as large as those for CeCoIn$_{5-x}$Sn$_x$ with the same $x$ range \cite{rf:Bauer2006,rf:Gofryk2012}. The difference in the magnitude of the impurity scattering between these doped alloys might be governed by the occupied impurity position rather than the ionic radius of impurity, because both the Ni and Sn atoms have neighbor atomic numbers of Co and In, respectively. From these considerations, we suggest that the impurities in the CeIn layer disturb the SC order parameter more effectively than those in the Co layer. Such an argument is also expected to be applicable to the difference in the rate $dT_{\rm c}/dx$ between these alloys. In general, the strength of the hybridization between the electrons in the impurity atoms and the f electrons in the Ce ions strongly depends on the position of the impurity, and it would be one of the reasons for the impurity-position dependence of the impurity scattering, and the values of $d(\Delta C_p/T_{\rm c})/d x$ and $dT_{\rm c}/d x$. On the other hand, the electrons added by the Ni and Sn impurities have different orbital angular momentum from each other, and this difference would also affect the discrepancies in $d(\Delta C_p/T_{\rm c})/d x$ and $dT_{\rm c}/d x$ through the c-f mixing. Photoemission-spectroscopy measurement is expected to provide more precise information concerning the role of the impurities on the electronic state in these doped alloys.

Another remarkable feature concerning the crystallographic position of the impurity is that the widths of the hysteresis loop in $M(B)$ for Ce(Co,Ni)In$_5$ are much smaller than those for CeCo(In,Zn)$_5$ with nearly the same amount of doped ions \cite{rf:Yoko2015}. We previously observed that the $c$-axis $M(B)$ curves in CeCo(In$_{1-x}$Zn$_x$)$_5$ have a very large hysteresis loop in the SC phase regardless of the value of $x$ ($x>0$). It is thus expected that this enlarged hysteresis loop is not mainly caused by the AFM ordering, but by the impurities set within the active CeIn layer. Namely, the doped impurities should contribute to the pinning of the flux perpendicular to this layer. On the other hand, such an effect would be limited in Ce(Co,Ni)In$_5$ when the SC order parameter is not tightly coupled with the electronic state in the Co layer. In this situation, the doped Ni impurities do not act as strong pinning centers, thereby yielding the small irreversible curves in $M(B)$ as observed in the present investigation. We suggest from the above two features that the coherence in the CeIn layers strongly affects the stability of the SC order. Indeed, the significance of the spatial coherence of the electronic state at around the impurity has been argued in a recent nuclear-quadrupole-resonance investigation of the Sn- and Cd-doped CeCoIn$_5$ \cite{rf:Sakai2015}.

\subsection{AFM quantum criticality in CeCo$_{1-x}$Ni$_x$In$_5$}
Here, we also discuss the nature of the NFL anomalies in CeCo$_{1-x}$Ni$_x$In$_5$. In pure CeCoIn$_5$, the existence of the AFM-QCP at $\sim H_{\rm c2}$ is inferred from the $-\ln T$ divergence in $C_p/T$ \cite{rf:Bianchi2003-2} and the weak $T$-linear increase in $M/B$ with decreasing temperature \cite{rf:Tayama2002}. On the other hand, our present investigation reveals that the $-\ln T$ divergence occurs in both $C_p/T$ and $M/B$ at $x=0.25$. In particular, the $-\ln T$ divergence in $M/B$ and its suppression by increasing magnetic field suggest that the NFL anomalies originate from the magnetic degrees of freedom with the AFM correlation \cite{rf:Yoko2016}. It is also expected from the comparison of the $M/B$ data between $x=0$ and $x=0.25$ that the NFL anomalies observed in these compounds have the same origin. As a consequence, it is considered that the Ni-doped alloys for $x \le 0.25$ are still located in the proximity of the AFM-QCP, and the further Ni substitution up to 30\% creates a distance from it because the $-\ln T$ dependence in $C_p/T$ becomes weak and the exponent $n$ in the electrical resistivity increases toward the value (=2) expected in the Fermi-liquid state. 

It is remarkable that the $-\ln T$ divergence in $M/B$ at $x=0.25$ has a stronger temperature dependence than the $T$-linear increase observed at $x=0$. The former temperature dependence in $M/B$ may be realized only at $B\sim 0$, although it is never detected for $x \le 0.20$ because the SC order always disturbs the observation of the NFL behavior for $B < \mu_0H_{\rm c2}$. Alternatively, such a difference in the $M/B$ curves may be caused by the reduction in dimensionality of the AFM quantum critical fluctuation inherently generated by doping Ni. We expect that a microscopic investigation such as a neutron scattering experiment can resolve this issue.

On the other hand, one may suspect that the disorder enhanced at $x\sim 0.25$ predominantly contributes to the NFL behavior at $x=0.25$, since the disorder effect would be unavoidable in the doped alloys \cite{rf:Miranda2005}. However, we consider that the disorder effect does not seem to be the main reason for the $-\ln T$ divergence in $M/B$ and $C_p/T$. If the energy distribution width of the low-lying electronic states due to the disorder is a few or ten Kelvin in accordance with the temperature range of the $-\ln T$ dependence in $M/B$, it is expected that the magnitude of $C_p/T$ in the corresponding temperature range would significantly depend on the strength of the disorder, and eventually on $x$. However, the observed $C_p/T$ values above $\sim 3\ {\rm K}$ are independent of $x$, in contrast to the above assumption. 

Despite the above considerations, no direct evidence of the origin of the NFL behavior has been obtained in this study. To clarify it, we plan to perform the precise measurements of $C_p$, $M$, and $\rho$ under magnetic fields. It is particularly remarkable that the $T/B$-scaling properties of the $C_p$ data related to the quantum critical fluctuation have been revealed in pure CeCoIn$_5$ \cite{rf:Bianchi2003-2}. Thus, the application of the similar analysis for the $C_p$ and $M$ data obtained from the Ni-doped alloys may provide a clue for resolving the origin of the NFL behavior presently revealed.

\section{Summary}
The thermal, magnetic, and transport properties of the SC and NFL states for CeCo$_{1-x}$Ni$_x$In$_5$ ($x \le 0.30$) alloys are presented for the first time. It is found that the SC order is monotonically suppressed by doping Ni into CeCoIn$_5$, and then replaced by the paramagnetic state involving the NFL behavior above $x=0.25$. These tendencies resemble those revealed in Sn- and Pt-doped CeCoIn$_5$ \cite{rf:Bauer2005,rf:Bauer2006,rf:Ramos2010,rf:Gofryk2012}, and hence, our present investigation adds another example for the reduction in the SC order parameter caused by the increase in the number of electron carriers by doping \cite{rf:Gofryk2012}. However, the present study has also revealed the quantitative differences in the trends of the SC suppression between Ce(Co,Ni)In$_5$ and the other alloys. In particular, we have compared the decreasing rate of the $\Delta C_p/T_{\rm c}$ values [$d(\Delta C_p/T_{\rm c})/d x$] and the magnitudes of the hysteresis loop in $M(B)$ between the Ni- and Sn-doped (or Zn-doped) alloys, and suggest that the coherence of the electronic state within the active CeIn layer is crucial for the stability of the SC order.

In addition, the present investigation has clarified that the NFL behavior at $x=0.25$ involves the $-\ln T$ divergence in both $C_p/T$ and $M/B$, and the $-\ln T$ function in $M/B$ changes into the nearly $T$-linear dependence by applying weak magnetic fields. Since the latter $T$ dependence coincides fairly well with that observed at $\sim H_{\rm c2}$ in pure CeCoIn$_5$, we consider that the compounds with both $x=0$ and $x=0.25$ are located in the proximity of the same AFM-QCP. 

\section*{Acknowledgment}
\begin{acknowledgment}
M.Y. is grateful to Y. Sakon, K. Suzuki, Y. Oshima, Y. Kono, and S. Kittaka for their support of the magnetization measurements, and to Y. Matsuda for helpful discussion. This work was carried out as a joint research in the Institute for Solid State Physics (ISSP), The University of Tokyo. The measurements at the ISSP were supported in part by Grants-in-Aid for Scientific Research on Innovative Areas ``J-Physics" (15H05883) from MEXT and KAKENHI (15H03682) from JSPS.
\end{acknowledgment}


\begin{thebibliography}{99}
\bibitem{rf:Mathur98}N. D. Mathur, F. M. Grosche, S. R. Julian, I. R. Walker, D. M. Freye, R. K. W. Haselwimmer, and G. G. Lonzarich, Nature {\bf 394}, 39 (1998).
\bibitem{rf:Lohneysen2007}H. v. L\"ohneysen, A. Rosch, M. Vojta, and P. W\"olfle, Rev. Mod. Phys. {\bf 79}, 1015 (2007).
\bibitem{rf:Pfleiderer2009} C. Pfleiderer, Rev. Mod. Phys. {\bf 81}, 1551 (2009).
\bibitem{rf:Stewart2001} G. R. Stewart, Rev. Mod. Phys. {\bf 73}, 797 (2001).
\bibitem{rf:Hall2001} D. Hall, E. C. Palm, T. P. Murphy, S. W. Tozer, Z. Fisk, U. Alver, R. G. Goodrich, J. L. Sarrao, P. G. Pagliuso, and T. Ebihara, Phys. Rev. B {\bf 64}, 212508 (2001).
\bibitem{rf:Petrovic2001} C. Petrovic, P. G. Pagliuso, M. F. Hundley, R. Movshovich, J. L. Sarrao, J. D. Thompson, Z. Fisk, and P. Monthoux, J. Phys.: Condens. Matter {\bf 13}, L337 (2001).
\bibitem{rf:Izawa2001} K. Izawa, H. Yamaguchi, Y. Matsuda, H. Shishido, R. Settai, and Y. Onuki, Phys. Rev. Lett. {\bf 87}, 057002 (2001).
\bibitem{rf:An2010} K. An, T. Sakakibara, R. Settai, Y. Onuki, M. Hiragi, M. Ichioka, and K. Machida, Phys. Rev. Lett. {\bf 104}, 037002 (2010).
\bibitem{rf:Park2008} W. K. Park, J. L. Sarrao, J. D. Thompson, and L. H. Greene, Phys. Rev. Lett. {\bf 100}, 177001 (2008).
\bibitem{rf:Tayama2002} T. Tayama, A. Harita, T. Sakakibara, Y. Haga, H. Shishido, R. Settai, and Y. Onuki, Phys. Rev. B {\bf 65}, 180504(R) (2002).
\bibitem{rf:Ikeda2001} S. Ikeda, H. Shishido, M. Nakashima, R. Settai, D. Aoki, Y. Haga, H. Harima, Y. Aoki, T. Namiki, H. Sato, and Y. Onuki, J. Phys. Soc. Jpn. {\bf 70}, 2248 (2001).
\bibitem{rf:Bianchi2002} A. Bianchi, R. Movshovich, N. Oeschler, P. Gegenwart, F. Steglich, J. D. Thompson, P. G. Pagliuso, and J. L. Sarrao, Phys. Rev. Lett. {\bf 89}, 137002 (2002).
\bibitem{rf:Bianchi2003-1} A. Bianchi, R. Movshovich, C. Capan, P. G. Pagliuso, and J. L. Sarrao, Phys. Rev. Lett. {\bf 91}, 187004 (2003).
\bibitem{rf:Radovan2003} H. A. Radovan, N. A. Fortune, T. P. Murphy, S. T. Hannahs, E. C. Palm, S. W. Tozer, and D. Hall, Nature {\bf 425}, 51 (2003).
\bibitem{rf:Kakuyanagi2005} K. Kakuyanagi, M. Saitoh, K. Kumagai, S. Takashima, M. Nohara, H. Takagi, and Y. Matsuda, Phys. Rev. Lett. {\bf 94}, 047602 (2005).
\bibitem{rf:Young2007} B.-L. Young, R. R. Urbano, N. J. Curro, J. D. Thompson, J. L. Sarrao, A. B. Vorontsov, and M. J. Graf, Phys. Rev. Lett. {\bf 98}, 036402 (2007).
\bibitem{rf:Kenzelmann2008} M. Kenzelmann, Th. Strassle, C. Niedermayer, M. Sigrist, B. Padmanabhan, M. Zolliker, A. D. Bianchi, R. Movshovich, E. D. Bauer, J. L. Sarrao, and J. D. Thompson, Science {\bf 321}, 1652 (2008). 
\bibitem{rf:Tokiwa2008} Y. Tokiwa, R. Movshovich, F. Ronning, E. D. Bauer, P. Papin, A. D. Bianchi, J. F. Rauscher, S. M. Kauzlarich, and Z. Fisk, Phys. Rev. Lett. {\bf 101}, 037001 (2008).
\bibitem{rf:Bianchi2003-2} A. Bianchi, R. Movshovich, I. Vekhter, P. G. Pagliuso, and J. L. Sarrao, Phys. Rev. Lett. {\bf 91}, 257001 (2003).
\bibitem{rf:Paglione2003} J. Paglione, M. A. Tanatar, D. G. Hawthorn, E. Boaknin, R. W. Hill, F. Ronning, M. Sutherland, L. Taillefer, C. Petrovic, and P. C. Canfield, Phys. Rev. Lett. {\bf 91}, 246405 (2003).
\bibitem{rf:Hu2008} R. Hu, Y. Lee, J. Hudis, V. F. Mitrovic, and C. Petrovic, Phys. Rev. B {\bf 77}, 165129 (2008).
\bibitem{rf:Raymond2014} S. Raymond, S. M. Ramos, D. Aoki, G. Knebel, V. P. Mineev, and G. Lapertot, J. Phys. Soc. Jpn. {\bf 83}, 013707 (2014).
\bibitem{rf:Zapf2001} V. S. Zapf, E. J. Freeman, E. D. Bauer, J. Petricka, C. Sirvent, N. A. Frederick, R. P. Dickey, and M. B. Maple, Phys. Rev. B {\bf 65}, 014506 (2001).
\bibitem{rf:Yoko2006} M. Yokoyama, H. Amitsuka, K. Matsuda, A. Gawase, N. Oyama, I. Kawasaki, K. Tenya, and H. Yoshizawa, J. Phys. Soc. Jpn. {\bf 75}, 103703 (2006). 
\bibitem{rf:Yoko2008} M. Yokoyama, N. Oyama, H. Amitsuka, S. Oinuma, I. Kawasaki, K. Tenya, M. Matsuura, K. Hirota, and T. J. Sato, Phys. Rev. B {\bf 77}, 224501 (2008).
\bibitem{rf:Ohira-Kawamura2007} S. Ohira-Kawamura, H. Shishido, A. Yoshida, R. Okazaki, H. Kawano-Furukawa, T. Shibauchi, H. Harima, and Y. Matsuda, Phys. Rev. B {\bf 76}, 132507 (2007).
\bibitem{rf:Pham2006} L. D. Pham, T. Park, S. Maquilon, J. D. Thompson, and Z. Fisk, Phys. Rev. Lett. {\bf 97}, 056404 (2006).
\bibitem{rf:Nicklas2007} M. Nicklas, O. Stockert, T. Park, K. Habicht, K. Kiefer, L. D. Pham, J. D. Thompson, Z. Fisk, and F. Steglich, Phys. Rev. B {\bf 76}, 052401 (2007).
\bibitem{rf:Yoko2014} M. Yokoyama, K. Fujimura, S. Ishikawa, M. Kimura, T. Hasegawa, I. Kawasaki, K. Tenya, Y. Kono, and T. Sakakibara, J. Phys. Soc. Jpn. {\bf 83}, 033706 (2014).
\bibitem{rf:Yoko2015}M. Yokoyama, H. Mashiko, R. Otaka, Y. Sakon, K. Fujimura, K. Tenya, A. Kondo, K. Kindo, Y. Ikeda, H. Yoshizawa, Y. Shimizu, Y. Kono, and T. Sakakibara, Phys. Rev. B {\bf 92}, 184509 (2015).
\bibitem{rf:Bauer2005} E. D. Bauer, C. Capan, F. Ronning, R. Movshovich, J. D. Thompson, and J. L. Sarrao, Phys. Rev. Lett. {\bf 94}, 047001 (2005).
\bibitem{rf:Bauer2006} E. D. Bauer, F. Ronning, C. Capan, M. J. Graf, D. Vandervelde, H. Q. Yuan, M. B. Salamon, D. J. Mixson, N. O. Moreno, S. R. Brown, J. D. Thompson, R. Movshovich, M. F. Hundley, J. L. Sarrao, P. G. Pagliuso, and S. M. Kauzlarich, Phys. Rev. B {\bf 73}, 245109 (2006).
\bibitem{rf:Ramos2010} S. M. Ramos, M. B. Fontes, E. N. Hering, M. A. Continentino, E. Baggio-Saitovich, F. Dinola Neto, E. M. Bittar, P. G. Pagliuso, E. D. Bauer, J. L. Sarrao, and J. D. Thompson, Phys. Rev. Lett. {\bf 105}, 126401 (2010).
\bibitem{rf:Daniel2005} M. Daniel, E. D. Bauer, S.-W. Han, C. H. Booth, A. L. Cornelius, P. G. Pagliuso, and J. L. Sarrao, Phys. Rev. Lett. {\bf 95}, 016406 (2005).
\bibitem{rf:Gofryk2012} K. Gofryk, F. Ronning, J.-X. Zhu, M. N. Ou, P. H. Tobash, S. S. Stoyko, X. Lu, A. Mar, T. Park, E. D. Bauer, J. D. Thompson, and Z. Fisk, Phys. Rev. Lett. {\bf 109}, 186402 (2012).
\bibitem{rf:Sidorov2002} V. A. Sidorov, M. Nicklas, P. G. Pagliuso, J. L. Sarrao, Y. Bang, A. V. Balatsky, and J. D. Thompson, Phys. Rev. Lett. {\bf 89}, 157004 (2002).
\bibitem{rf:Sakakibara94} T. Sakakibara, H. Mitamura, T. Tayama, and H. Amitsuka, Jpn. J. Appl. Phys. {\bf 33}, 5067 (1994).
\bibitem{rf:Paglione2007} J. Paglione, T. A. Sayles, P.-C. Ho, J. R. Jeffries, and M. B. Maple, Nat. Phys. {\bf 3}, 703 (2007).
\bibitem{rf:Werthammer66} N. R. Werthammer, E. Hefland, and P. C. Hohenberg, Phys. Rev. {\bf 147}, 295 (1966).
\bibitem{rf:Pt-and-Ni} For instance, the discrepancies in the $x$ variations of $T_{\rm c}$ and $T_{\rm max}$ between Ni- and Pt-doped CeCoIn$_5$ would arise from the differences in the c-f hybridization strength and the chemical pressure depending on the element of impurity, since the period and the ionic radius of the Ni and Pt atoms are different from each other, whereas both the Ni and Pt ions occupy identical crystallographic positions and nominally add an electron to the crystal for each atom. 
\bibitem{rf:Hudson2001} E. W. Hudson, K. M. Lang, V. Madhavan, S. H. Pan, H. Eisaki, S. Uchida, and J. C. Davis, Nature {\bf 411}, 920 (2001).
\bibitem{rf:Sakai2015} H. Sakai, F. Ronning, J.-X. Zhu, N. Wakeham, H. Yasuoka, Y. Tokunaga, S. Kambe, E. D. Bauer, and J. D. Thompson, Phys. Rev. B {\bf 92}, 121105 (2015).
\bibitem{rf:Yoko2016} To our best knowledge, the $-\ln T$ divergence in the magnetic susceptibility is observed for the first time in CeCoIn$_5$ and its doped alloys with the dense-limit Ce concentration.
\bibitem{rf:Miranda2005} E. Miranda and V. Dobrosavljevi\'c, Rep. Prog. Phys. {\bf 68}, 2337 (2005).
\end{thebibliography}
\end{document}